\documentclass[osajnl,twocolumn,showpacs,superscriptaddress,10pt]{revtex4-1} 
\usepackage{amsmath,amssymb,graphicx}

\begin{document}

\title{Tracking surface plasmon pulses using ultrafast leakage imaging}

\author{Yuri Gorodetski}
\affiliation{Electrical Engineering and Electronics Department, Ariel University, 40700, Ariel, Israel}
\author{Thibault Chervy}
\author{Shaojun Wang}
\author{James A. Hutchison} 
\affiliation{ISIS \& icFRC, Universit\'{e} de Strasbourg and CNRS (UMR 7006), 8 all\'{e}e Gaspard Monge, 67000 Strasbourg, France}
\author{Aur\'elien Drezet}
\affiliation{Institut N\'eel, CNRS and Universit\'e Joseph-Fourier (UPR 2940), 25 rue des Martyrs, 38042 Grenoble, France}
\author{Cyriaque Genet}\email{Corresponding author: genet@unistra.fr}
\author{Thomas W. Ebbesen}
\affiliation{ISIS \& icFRC, Universit\'{e} de Strasbourg and CNRS (UMR 7006), 8 all\'{e}e Gaspard Monge, 67000 Strasbourg, France}

\begin{abstract}
We introduce a new method for performing ultrafast imaging and tracking of surface plasmon wave packets that propagate on metal films. We demonstrate the efficiency of leakage radiation microscopy implemented in the time domain for measuring both group and phase velocities of near-field pulses with a high level of precision. The versatility of our far-field imaging method is particularly appealing in the context of ultrafast near-field optics.
\end{abstract}

\maketitle 

Among the different surface plasmon imaging techniques, leakage radiation (LR) microscopy is a powerful method for imaging surface plasmon (SP) modes propagating on metal-dielectric interfaces \cite{Hecht,Bouhelier,Drezet,Hohenau,DrezetPRL2013}. This imaging method has been implemented in a great variety of situations in SP optics, ranging from surface plasmon circuitry \cite{BaudrionOptX2008,SteinPRL2010,MinovichPRL2011,LiPRL2011} to near-field weak measurements \cite{GorodetskiPRL2012}, both at the classical and quantum levels \cite{CucheNanoLett2010}. Recently, this technique has been combined with interference microscopy, providing not only the amplitude but also the phase of the leakage signal \cite{HerzigOptLett2013}. 

In this Article, we operate LR microscopy in the time-domain and demonstrate its efficiency for performing ultrafast imaging of propagating SP wave packets at the diffraction limit. While our scheme leads to the simultaneous measurement of both the group and the phase velocities of the SP wave packet, it also provides a unique method to resolve higher-order dispersive effects associated with the plasmonic signal, such as plasmonic group velocity dispersion effects. The simplicity of our all-optical method, that does not involve any raster-scanned local probe nor non-linear detection processes, presents a clear advantage with respect to the sophistication of recently proposed near-field pulse tracking techniques (such as phase-sensitive time-resolved photon scanning tunneling microscopy \cite{BalistreriScience2001,GersenPRE2003}, time-resolved two-photon photoemission electron microscopy \cite{KuboNanoLett2005,LemkeNanoLett2013,GongNanoLett2015} or pulse-tracking via far-field SP scattering interference imaging \cite{RokitskiPRL2005}).

\begin{figure}[htbp]
\centering
{\includegraphics[width=\linewidth]{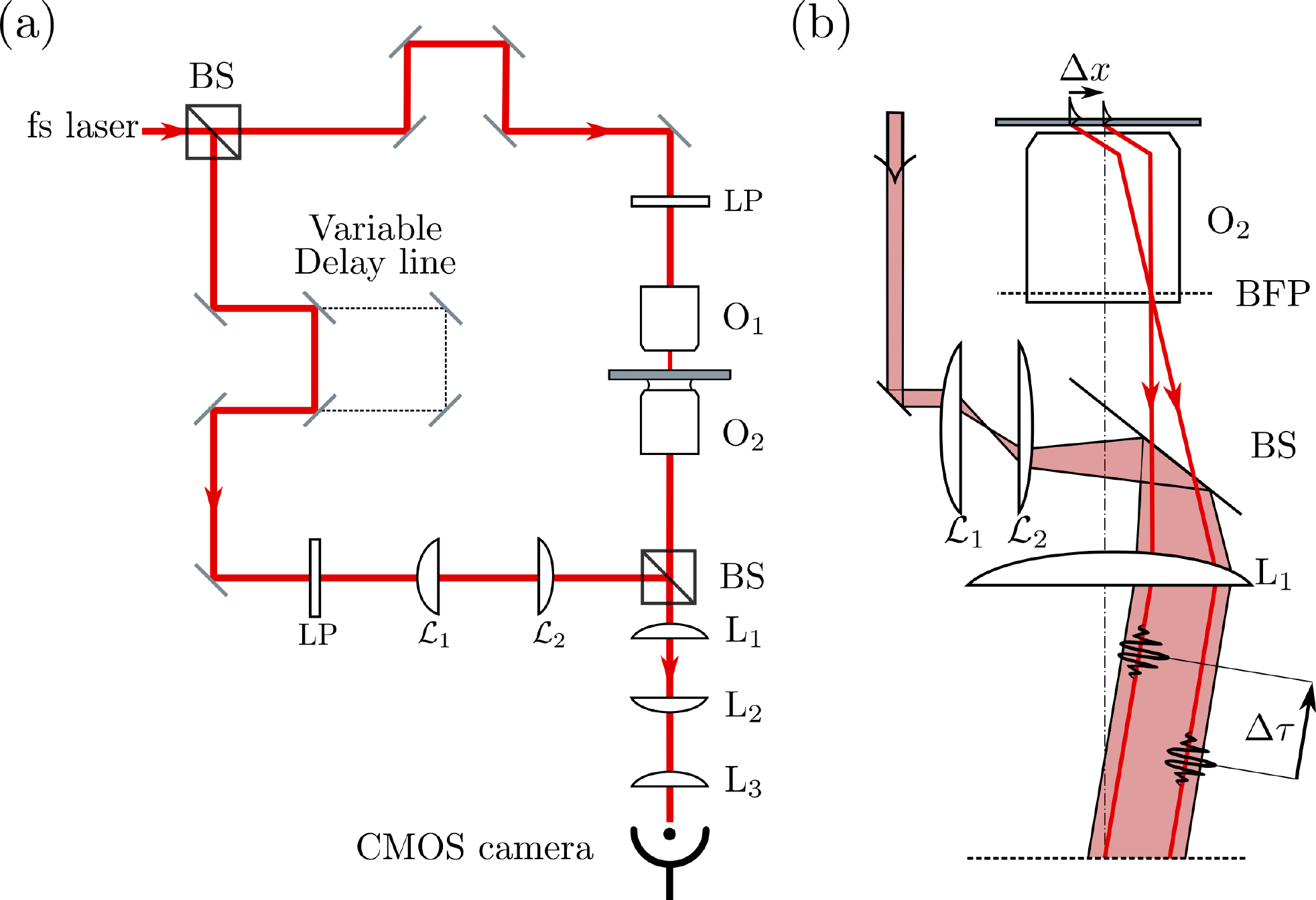}}
\caption{(a) Schematics of the LR interferometer. The fs pulsed laser beam ($\lambda_0=800$ nm, pulse duration $\tau_0=120$ fs) generated by a Ti:Sapphire oscillator ($1$ kHz) is split into two arms using a non-polarizing beam splitter (BS). One arm is sent through a leakage radiation microscope made of an illumination objective (O$_1$, magnification $20\times$ and numerical aperture ${\rm NA}=0.45$) that excites an SP wave packet on a thin $70$ nm metal (Au) film sputtered on a glass substrate. The LR signal is collected with a high numerical aperture collection objective (O$_2$, magnification $100\times$, ${\rm NA}=1.3$). The second arm (reference arm) is optically delayed (time-resolution of $10$ fs for a total range of $80$ ps) and recombined with the leakage signal. The $\mathcal{L}_1,\mathcal{L}_2$ telescope is crucial in order to ensure proper beam overlap and in order to match the curvatures between the two beams, as explained in the main text. After the recombination on the second beam splitter (BS), the interferogram is imaged using a sequence of lenses (performing both real and Fourier space imaging) on a CMOS camera. The linear polarization between the two beams is controlled by a series of polarizers and half-wave plates on each arm of the interferometer represented by the LP polarization stage. (b) Detailed view of the LR microscope with the SP field shown leaking through the metal film in the glass substrate with an angle $\alpha$ fixed by the SP wave vector $k_{\rm SP}$ as $\sin\alpha= {\rm Re}[k_{\rm SP}] / (n_g k_0)$, with $n_g$ the refractive index of glass and $k_0$ the vacuum wave number. The recombination optics detailed after the back focal plane (BFP) of the objective O$_2$ shows the delay $\Delta\tau$ induced by the SP mode propagating over $\Delta x$ on the metal film. The bottom dotted horizontal line corresponds to the image plane of the objective where successive interferograms are imaged, as in Fig.\ref{fig4}.}
\label{fig1}
\end{figure}

\begin{figure}[htbp]
\centering
{\includegraphics[width=6cm]{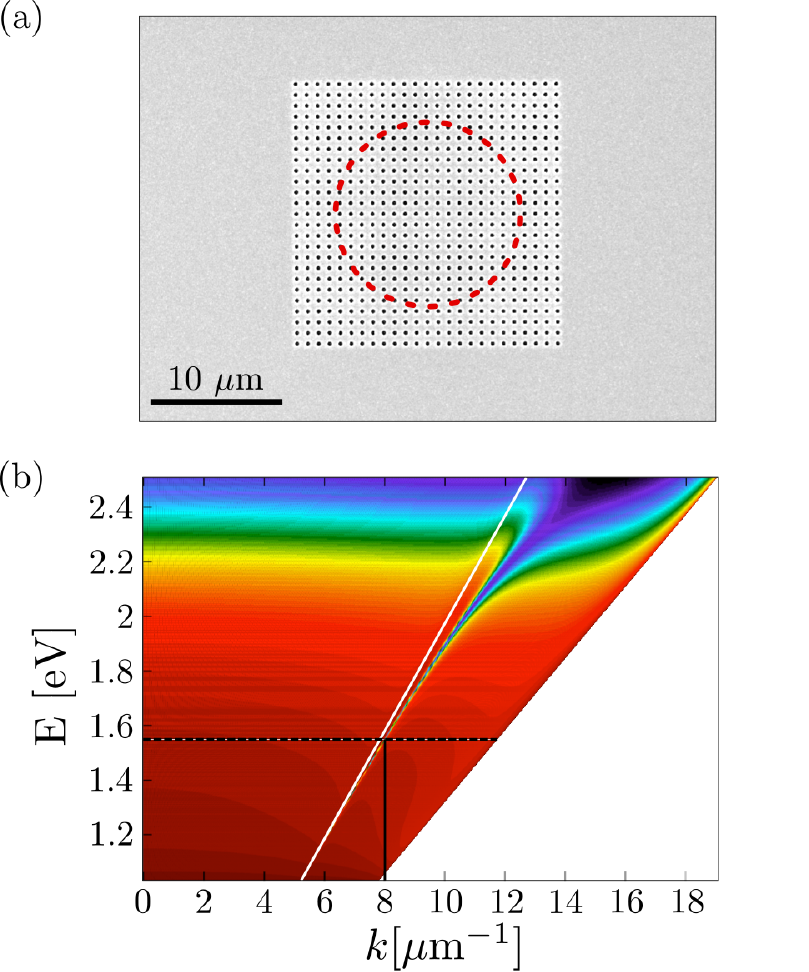}}
 \caption{(a) Scanning Electron Microscope image of the launching hole array. The array, milled through a thin Au film, is a square array of period $p=790$ nm and hole diameter $d=100$ nm. The superimposed dashed circle corresponds to the typical spot size of the excitation beam when defocused on the array. (b) Color-coded reflectivity calculated with a T-matrix method for a thin ($70$ nm) Au metal slab as a function of the light angular frequency $\omega$ and the in-plane light wave vector  $k$. The evolution of the reflection minimum in the ($k,\omega$) plane draws the dispersion relation $k_{\rm SP} (\omega)$ for an SP mode propagating at the air-metal interface. The horizontal dashed line is positioned at the carrier frequency $\omega_0$ and the tilted white line represents the light cone $\omega = ck$ in air. Theoretical values for the SP propagation constant $\kappa (\omega_0)^{disp.}$ (indicated by the continuous vertical black line), SP loss $\Gamma$ and group velocity $(\partial \kappa / \partial \omega)_{\omega_0}^{disp.}$ are extracted from this dispersion relation.}
\label{fig2}
\end{figure}

Our experimental scheme, described in details in Fig. \ref{fig1}, consists in inserting an LR microscope within a Mach-Zehnder interferometer at the input of which a transform-limited laser pulse is evenly split into two beams. In one arm, the beam resonantly launches an SP wave packet on a thin metal film using an $(x,y)$ square hole array (shown in Fig. \ref{fig2} (a)) properly designed  and milled through the film. This beam is linearly polarized in the $x$ direction of the array, corresponding to the propagation direction of the SP wave packet. A high numerical aperture (NA) oil-immersion objective collects the plasmonic LR signal as a pulse $E_L$ that propagates along the arm of the interferometer. This $E_L$ pulse is then combined with the reference pulse $E_R$ coming from the other arm of the interferometer that can be time-delayed with respect to $E_L$ using a motorized optical delay line. The $E_R$ pulse is linearly polarized in the same direction as $E_L$ so that both pulses can interfere at the output of the interferometer. When the two pulses overlap in time and space, an interferogram is formed in the image plane of the collection objective. The central point of our scheme, sketched in Fig. \ref{fig1} (b), consists in monitoring interferograms as a function of the delay time between the two pulses. As we show below, this allows one to track SP wave packet propagation since successive interferograms record the evolution of the phase of the LR signal emitted by the SP at different positions along the metal film.

\begin{figure}[htbp]
\centering
{\includegraphics[width=6cm]{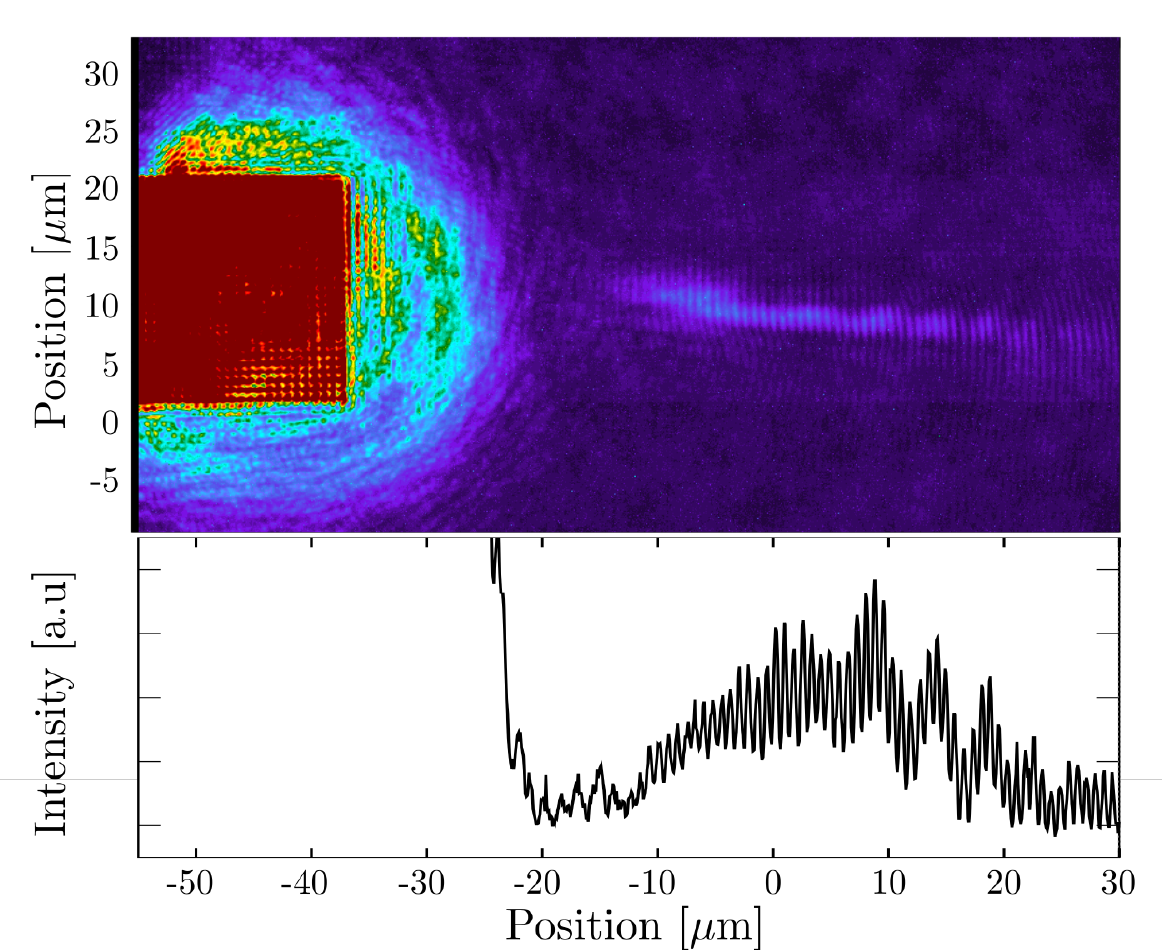}}
 \caption{(a) LR image recorded on the CMOS camera at the output of the interferometer for a time-delay $\tau=230 fs$ (see Fig. \ref{fig1}). While the light transmitted directly through the launching array saturates the left-hand side of the image frame, interference fringes are clearly observed away from the array within the time-averaged SP intensity distribution $|E_L|^2$ (image scale in $\mu{\rm m}$). (b) Intensity cross-cut performed from the middle of the array.}
\label{fig3}
\end{figure}

\begin{figure}[htbp]
\centering
{\includegraphics[width=7cm]{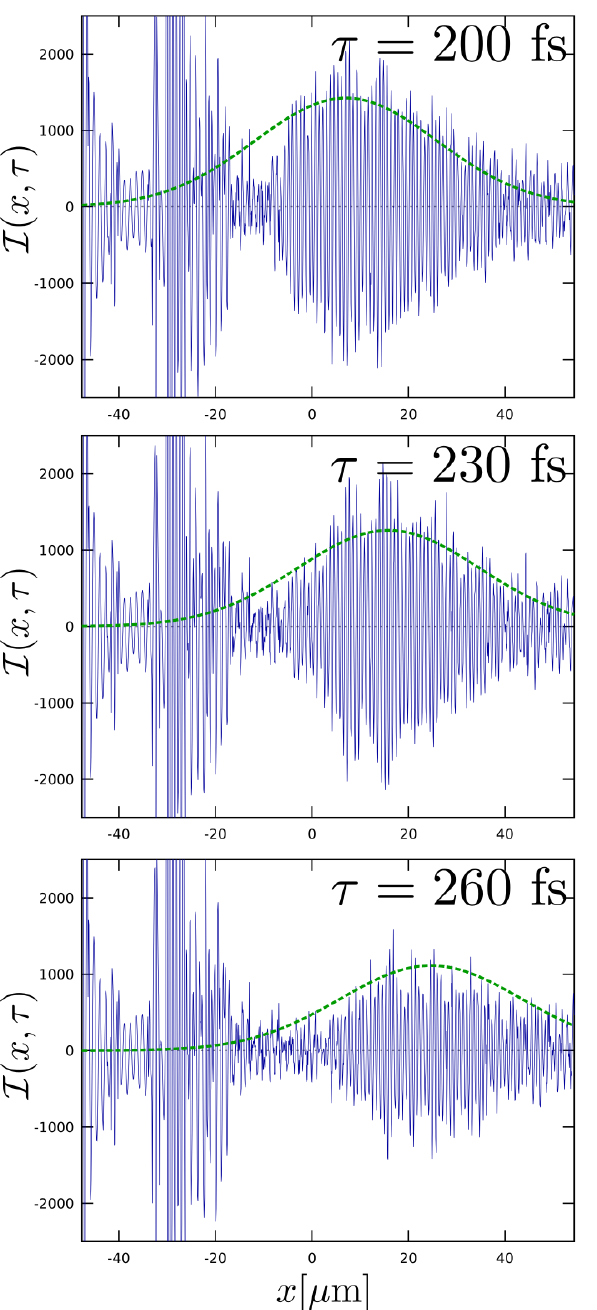}}
 \caption{Sequence of experimental time-resolved interferograms obtained by post-filtering (see main text) cross-cut images similar to Fig. \ref{fig3} (b) for time delays $\tau=200, 230, 260$ fs. These experimental interferograms are compared to the model ones given by Eq. (\ref{eqn:1}). The fitted space-time evolutions of the pulse envelope are displayed as dashed curves -see main text for the fitting parameters.  }
\label{fig4}
\end{figure}

An imaged interferogram is displayed in Fig. \ref{fig3} (a) for a given time-delay $\tau$ between $E_L$ and $E_R$. While the SP wave packet launched by the array is clearly observed, interference fringes are also seen away from the array (see image cross-cut in Fig. \ref{fig3} (b)) exactly where the two $E_L$ and $E_R$ pulses overlap. We emphasize that while the time-delay $\tau$ controls the time overlap between the pulses at the output of the interferometer, a spatial overlap between the moving pulse $E_L$ and the reference pulse $E_R$ must simultaneously be ensured at all times. This is done by expanding optically $E_R$ with a telescope, as drawn in Fig. \ref{fig1} (b). Moreover, this $\mathcal{L}_1,\mathcal{L}_2$ telescope gives the possibility to adjust the radii of curvature of both beams at the recombination plane, thus preventing any curvature phase mismatch from degrading the interferogram, as discussed further down. In such conditions, the time evolution of successive interferograms can be easily monitored, as shown in Fig. \ref{fig4} for interferograms separated from each other by a $\Delta \tau=30$ fs time delay. These interferograms have been obtained by cross-cutting the original real-space images (as done in Fig. \ref{fig3} (b)) after they have been post-filtered from the non-oscillatory time-averaged leakage contribution $|E_L|^2$ and from the direct transmission through the array. 

In order to describe the optical response of our system and to understand how the properties of the SP wave packet can be extracted from such interferograms, we model the plasmonic signal in the most simple way with the spatial evolution of an one-dimensional SP mode $\mathcal{P}(x)=e^{i\kappa |x|}e^{-\Gamma |x|/2}$, propagating on the metal film from both sides of the launching array, with a propagation constant $\kappa = {\rm Re}[k_{\rm SP}]$ given by the real part of the SP wave vector $k_{\rm SP}$ at the air-Au interface and an inverse decay length $ \Gamma=2\times{\rm Im}[k_{\rm SP}]$ determined on the SP loss. Such a 1D model nicely fits the spatial evolution of our experimental SP signals whose in-plane diffraction is strongly reduced by using a defocused excitation beam on the launching array \cite{SteinOL2012}.

Both propagation constant $\kappa$ and inverse decay length $\Gamma$ are directly determined respectively from the position and width of the reflectivity resonance calculated in Fig. \ref{fig2} (b) with the finite thickness of the metal film accounted for. Of course, the leakage signal $\mathcal{L}(x)$ decoupled in the far field by the collection objective is the convolution between the SP field $\mathcal{P}(x)$ and the point spread function of the collection objective accounting for its finite NA \cite{DrezetEPL2006}. But choosing a sufficiently high NA with ${\rm  NA}\cdot k_0 > {\rm Re}[k_{\rm SP}]$, the whole SP field can be imaged and we can therefore take $\mathcal{L}(x)\sim \mathcal{P}(x)$.

Within this framework, the interferograms can be directly evaluated considering an initial Gaussian pulse $E_0[\omega]=\exp{[-(\omega-\omega_0)^2/4\sigma_0^2]}$, with a carrier frequency $\omega_0$ corresponding to the laser central wavelength $\lambda_0=800$ nm and a transform-limited pulse width $\sigma_0\cdot\tau_0=\sqrt{2\ln 2}$. The reference ($R$) and leakage ($L$) pulses can be written as
\begin{equation}\label{eqn:1}
\begin{array}{l}
\displaystyle \textit{E}_{R} (z,\omega)\propto E_0[\omega] e^{ik_0 z} \\
\displaystyle \textit{E}_{L} (x,z,\omega)\propto E_0[\omega] \mathcal{P}(x) e^{ik_0 z} ,
\end{array} 
\end{equation}
where the free propagation in each arm of the interferometer is accounted for by the phase $ik_0 z$. 

Because the response time of the CMOS detector is much longer than the pulse duration $\tau_0$, the interference pattern recorded at the output of the interferometer is in fact time-averaged over a whole train of pulses, all of which are assumed to be identical. From the Wiener-Khintchine theorem, the interferogram can then be directly evaluated from the cross-correlation of the two $(L,R)$ signals as
\begin{eqnarray}
\label{eqn:1}
 \mathcal{I} (x,\tau) \propto \int\limits_{-\infty}^{\infty} {\rm d}\omega \left( e^{-i\omega \tau}{E}_{L} (x,z,\omega){E}_{R} ^\ast (z,\omega) + c.c.\right)
 \end{eqnarray}
where $c.c.$ stands for complex conjugation. 

By expanding up to first-order the SP phase about its value at the pulse carrier frequency as $\kappa(\omega)\sim \kappa (\omega_0)+(\partial \kappa /\partial\omega)_{\omega_0}\cdot (\omega-\omega_0)$, the interferogram is calculated as
\begin{eqnarray}
\label{eqn:2}
 \mathcal{I} (x,\tau) \propto e^{-\frac{\Gamma}{2} |x|} e^{-{\sigma_0^2}\left(\tau - \tau_g\right)^2} \cos \left[\kappa (\omega_0)\cdot |x|+\omega_0\tau\right].
 \end{eqnarray}
As seen in this expression, the phase $e^{i\kappa|x|}$ added by the plasmonic contribution actually corresponds to a spatial-heterodyne interference between the two pulses. 

Within the plasmonic decay length, $\mathcal{I} (x,\tau)$ is therefore characterized by a Gaussian envelope evolving with a group delay time $\tau_g=(\partial \kappa /\partial\omega)_{\omega_0}\cdot |x|$, corresponding to an SP pulse propagating on the film with a group velocity $v_g(\omega_0)=(\partial \omega /\partial\kappa)_{\omega_0}$. The interferogram is also characterized by a carrier signal  with a spatial carrier frequency $\kappa (\omega_0)$ that corresponds to the fringes observed experimentally in Fig. \ref{fig4}. The SP phase velocity is directly derived from this carrier signal as $v_\phi = \omega_0 / \kappa (\omega_0)$. 

\begin{figure}[htbp]
\centering
{\includegraphics[width=8cm]{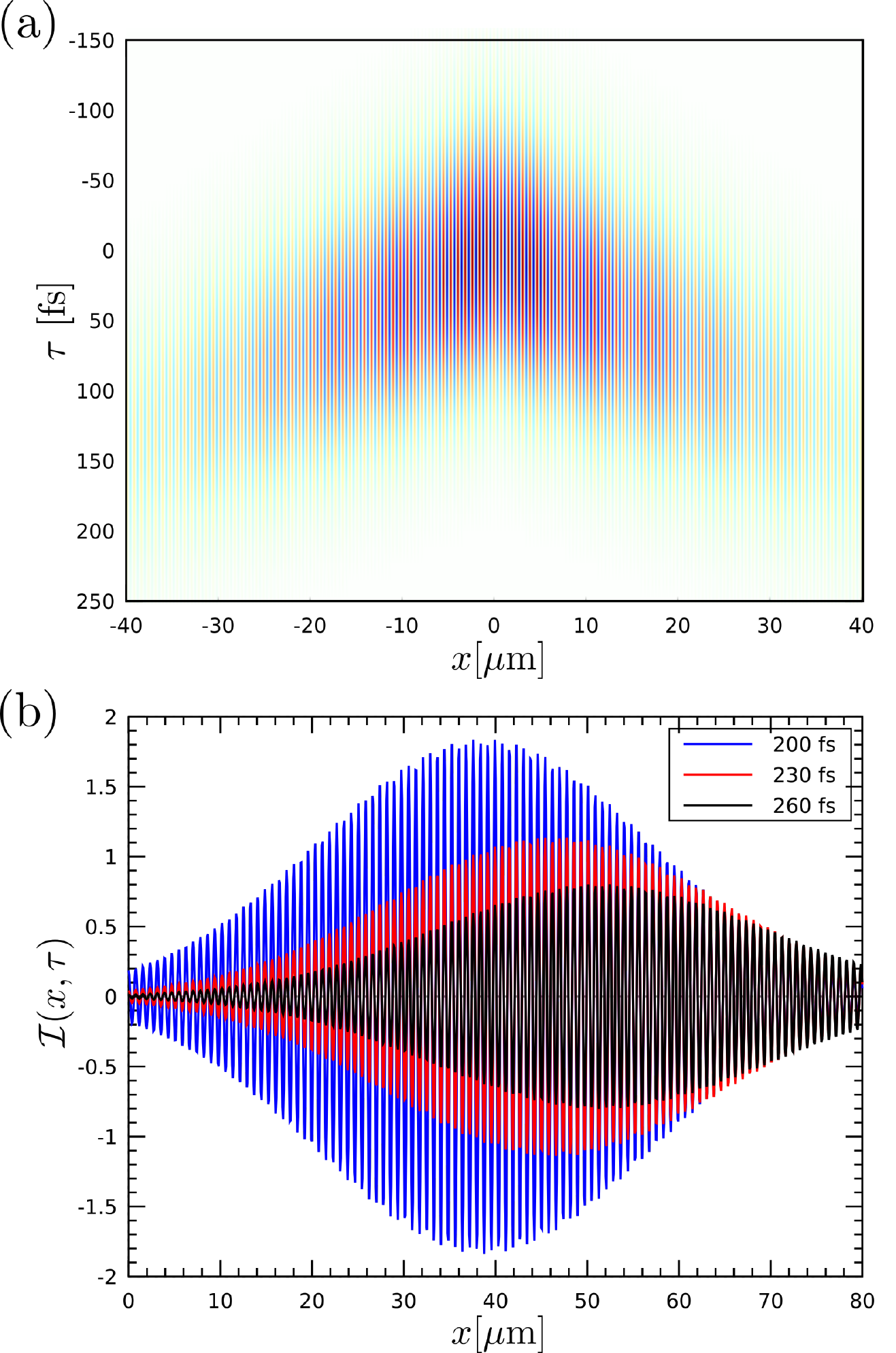}}
 \caption{(a) Calculated interferogram $\mathcal{I} (x,\tau)$ using the values of the group and phase velocities extracted from the dispersion relation of Fig. \ref{fig4}. The SP damping rate is taken from the Lorentzian modeling of the LR Fourier space with $\Gamma = 0.024~\mu$m$^{-1}$. (b) Cross-sections along the positive $x$ axis for the three successive delay times chosen as in Fig. \ref{fig3}.}
\label{fig5}
\end{figure}

Around the pulse carrier frequency $\omega_0$, the SP dispersion relation associated with the resonance profile of the reflectivity in Fig. \ref{fig2} (b) gives the expected SP propagation constant $\kappa (\omega_0)^{disp.}=8.025~\mu{\rm m}^{-1}$ and group velocity $v_g(\omega_0)^{disp.}=2.934\cdot10^8~{\rm m}\cdot{\rm s}^{-1}$. Using these values, the interferogram $\mathcal{I} (x,\tau)$ is calculated and drawn in the $(\tau,x)$ space-time diagram of Fig. \ref{fig5}(a), considering the initial $E_0[\omega]$ pulse for the SP excitation beam. The time evolution of $\mathcal{I} (x,\tau)$ is also shown in Fig. \ref{fig5}(b) as cross-sections taken at three successive time delays $\Delta \tau=30$ fs.

In evaluating Eq. (\ref{eqn:2}), we have neglected any source of group velocity dispersion (GVD) that could affect $\mathcal{I} (x,\tau)$. It is clear that as pulsed signals, $L$ and $R$ beams both experience GVD while propagating through the series of the optical elements of the interferometer. For interferograms of co-propagating pulses recorded in an image plane at a fixed position along the $z$-axis, GVD is not expected to have any influence apart from some pulse broadening and fixed phase offset on the interference carrier signal. But the situation is different in our spatial-heterodyne setup. The coupling of the excitation beam into an SP wave packet is a coherent process and implies that the GVD of the excitation beam is transferred to the SP beam which thereby becomes chirped, independently of the actual dispersive characteristics of the SP itself. In our experiment however, considering our rather long pulse width and the weakly dispersive optics involved in the setup, this GVD effect can be safely neglected \cite{footnote2}.

We have also neglected the effect on $\mathcal{I} (x,\tau)$ of plasmonic GVD itself. This is fully justified given the practically linear $(\kappa,\omega)$ relation around the pulse carrier frequency $\omega_0$ calculated from the reflectivity profile of Fig. \ref{fig2} (b). Nevertheless, we emphasize that it is straightforward to account for such a second-order dispersion effect with our scheme, providing the capacity to probe not only the linear evolution of the dispersion relation of the SP wave packet but also its local curvature (see \cite{footnote1}). 

We now compare the measured interferograms displayed in Fig. \ref{fig4} to the calculated ones plotted in Fig. \ref{fig5}. Using a simple fitting procedure, this comparison allows the measurement of the experimental values for the group and phase velocities of the SP wave packet. Starting with a transform-limited pulse width $\sigma_0$, the SP damping rate $\Gamma = 0.024~\mu$m$^{-1}$ and the SP group velocity $v_g^{disp.}$, and taking $A_0$ as a fixed amplitude of the interferogram, an initial fit is done on the first interferogram that sets the initial $(x_0,\tau_0)$ space-time coordinates. From these coordinates, the next $\Delta\tau=30$ fs interferogram is fit with the group velocity as the sole free parameter, keeping the $\sigma_0,\Gamma,A_0$ parameters as initially fixed. The results of the fits for each interferogram are shown as dashed envelopes in Fig. \ref{fig3}. Eventually and as it should, $v_g$ turns out to be fitted with a constant value through the whole fitting sequence over the successive time delays displayed in Fig. \ref{fig3}. As expected, this fitted value $v_g=2.901\cdot10^8~ {\rm m}\cdot{\rm s}^{-1}$ is very close to the calculated one $v_g^{disp.}$. 

\begin{figure}[htbp]
\centering
{\includegraphics[width=8cm]{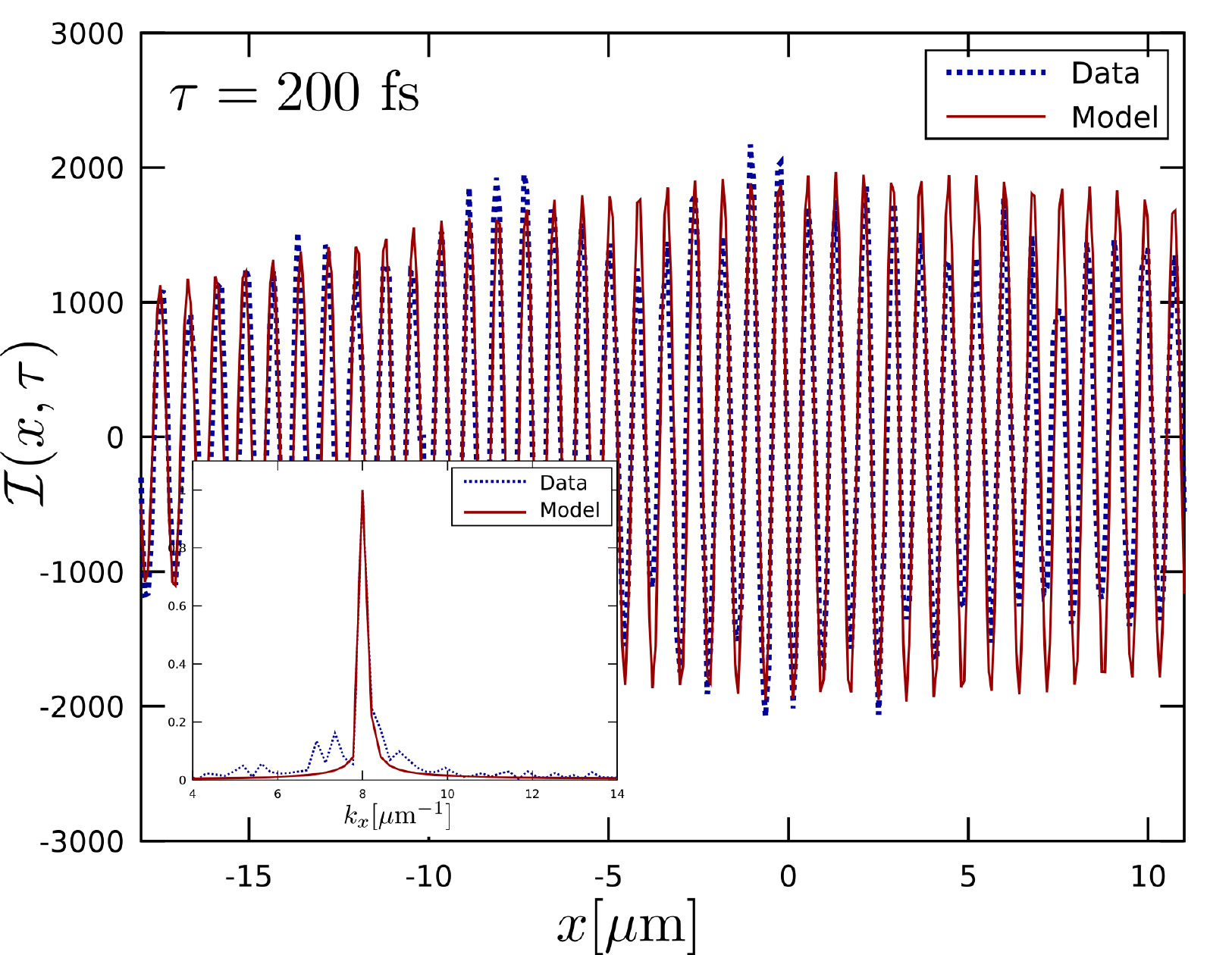}}
 \caption{Zoomed-in experimental interference fringes as observed in Fig. \ref{fig3} superimposed on the carrier signal calculated from Eq. (\ref{eqn:2}) using the fitting procedure described in the text. The inset displays the Fourier transforms of both the interference and the carrier signal. The two transforms overlap at the expected value of the SP propagation constant $\kappa^{disp.} (\omega_0)$.}
\label{fig6}
\end{figure}

The phase velocity can also be extracted by measuring the periodicity of the interference fringes in the experimental interferograms zoomed-in in Fig. \ref{fig6}. A Fourier transform of the image cross-section (inset in Fig. \ref{fig6}) allows the determination of the associated spatial frequency which should, according to Eq. (\ref{eqn:2}), be associated with the propagation constant $\kappa (\omega_0)$ of the SP wave packet at the central frequency $\omega_0$ of the pulse. At this stage, matching the $L$ and $R$ beam curvatures is particularly critical since any phase mismatch coming from differences in beam curvatures would prohibit the interferences in $\mathcal{I} (x,\tau)$ from being properly resolved. When this is done by adjusting carefully the $\mathcal{L}_1,\mathcal{L}_2$ telescope described in Fig. \ref{fig1} (b), the spatial frequency is measured with a high precision. The obtained value of $\kappa (\omega_0)\sim 8.02~\mu{\rm m}^{-1}$, corresponding to a phase velocity of $v_\phi\sim\left(2.94\pm0.01\right)\cdot10^8~{\rm m}\cdot{\rm s}^{-1}$, is very close to the one calculated from the dispersion relation at $\omega_0$. This is clearly seen in Fig. \ref{fig6} from the almost-perfect overlap between the experimental fringes and the  carrier signal calculated from Eq. (\ref{eqn:2}) using $v_\phi ^{disp.} = \omega_0 / \kappa (\omega_0)^{disp.}=2.936\cdot10^8~{\rm m}\cdot{\rm s}^{-1}$. These remarkable agreements both for the group and the phase velocities all confirm the ability of our setup to work as a high-resolution tracking method for resolving SP wave packets on the fs time-scale. 

To summarize, we have presented a new method involving LR microscopy that performs ultrafast imaging of SP pulses. Spatially heterodyning the leakage signal with a time-delayed reference pulse allows one to image the SP pulse propagation directly, with fs resolution and to determine the properties of the SP wave packet, in particular its group and phase velocities. The remarkable agreement between the observed interferograms and our modeling clearly demonstrates the potential of this all-optical scheme as a new tool in the field of SP optics. While the spatial resolution of the LR microscope is limited with respect to SNOM-based techniques, the unique level of control available on an LR microscope providing access to SP dynamics both in real and Fourier spaces with full polarization control, is particularly appealing. This possibility to combine time-resolved plasmonic imaging/tracking with polarization control (both in preparation and analysis sequences) opens very interesting perspectives, most obviously in the context of ultrafast signal processing in plasmonic media \cite{MacDonaldNatPhot2009,PohlPRB2012}.

We acknowledge support from the French Agence Nationale de la Recherche (ANR) Equipex Union (ANR-10-EQPX-52-01).


\begin{thebibliography}{1}
\bibitem{footnote2} The effect of GVD acquired by a pulse propagating over $\ell$ through a dispersive medium can measured by the pulse broadening factor $\sqrt{1+(2\mathcal{G}\sigma_0^{2}\ell)^2}$ where $\mathcal{G}$ is the GVD parameter of medium (in our case, optical glass BK7) and the pulse width $\sigma_0$. With $\mathcal{G}_{\rm BK7}\sim 50$ fs$^{2}\cdot$mm$^{-1}$ and $\tau_0=120$ fs, this ratio is larger than one by only ca. $10^{-5}\cdot$mm$^{-2}$.
\bibitem{footnote1} SP group velocity dispersion is accounted for by adding to the SP phase expansion a second-order term $\mathcal{G}=(\partial^2\kappa / \partial \omega^2)_{\omega_0}\cdot (\omega-\omega_0)^2$. The interferogram now writes as 
\begin{widetext}
\begin{eqnarray}
\mathcal{I} (x,\tau) &\propto& e^{-\frac{\Gamma |x|}{ 2}}\int\limits_{-\infty}^{\infty} {\rm d}\omega \ e^{-\frac{(\omega-\omega_0)^2 }{2\sigma_0^2}}\cos \left[\kappa (\omega_0)\cdot |x|+\omega_0 \tau -  (\tau -\tau_g)\cdot (\omega-\omega_0)+\frac{\mathcal{G}\cdot |x|}{2} \cdot (\omega-\omega_0)^2 \right]  \nonumber
\end{eqnarray}
\end{widetext}
which can be easily evaluated analytically.
\end{thebibliography}
\end{document}